\pdfoutput=1
\documentclass{article}
\usepackage[utf8]{inputenc}

\usepackage{setspace}
\usepackage{amsmath}
\usepackage{amssymb}
\usepackage{physics}
\usepackage[dvipsnames]{xcolor}
\usepackage{color}
\usepackage{tensor}
\usepackage{graphicx}
\usepackage{float}
\usepackage{xcolor, colortbl}
\usepackage{hyperref}
\usepackage{authblk}
\usepackage{mathtools}
\usepackage{hyperref}

\newcommand{\sltr}{\mathrm{SL}(2,\mathbb{R})}
\newcommand{\sltra}{\mathfrak{sl}(2,\mathbb{R})}
\newcommand{\kbad}{\kappa_\text{sing}}

\newenvironment{eqaed}
    {
    \begin{equation}
        \begin{aligned}
    }
    {
        \end{aligned}
    \end{equation}\ignorespacesafterend
    }

\newcommand{\tablespacing}{{\vskip 1em}}

\title{On Critical Exponents for Self-Similar Collapse}
\author[1]{Riccardo Antonelli}
\author[1]{Ehsan Hatefi}

\affil[1]{Scuola Normale Superiore and I.N.F.N,\protect\\ 
Piazza dei Cavalieri 7, 56126, Pisa, Italy\vspace{1em}}
\affil[ ]{\small\textit{riccardo.antonelli@sns.it, ehsan.hatefi@sns.it}}

\begin{document}

\maketitle

\begin{abstract}
    \noindent We explore systematically perturbations of self-similar solutions to the Einstein-axion-dilaton system, whose dynamics are invariant under spacetime dilations combined with internal $\sltra$ transformations. The self-similar solutions capture the enticing behavior ``critical'' systems on the verge of gravitational collapse, in arbitrary spacetime dimensions. Our methods rest on a combination of analytical and numerical tools, apply to all three conjugacy classes of $\sltra$ transformations and allow accurate estimates of the corresponding Choptuik exponents. It is well known that these exponents depend on the spacetime dimension and on the matter content. Our main result is that they also attain different values, even within a given conjugacy class, for the distinct types of critical solutions that we recently identified in the Einstein-axion-dilaton system.
\end{abstract}

\section{Introduction}

The study of spherically-symmetric collapse of scalar fields started in~\cite{Christodolou} and the emergence of a critical behavior in gravitational collapse was first considered by Choptuik in~\cite{Chop}, who showed that solutions on the verge of gravitational collapse feature some kind of spacetime self-similarity under special dilations. In addition, near-critical solutions display suggestive scaling laws. The example of~\cite{Chop} dealt with a minimally-coupled real scalar with initial conditions modulated by a parameter $p$ related to the field amplitude. If the choice $p=p_\text{crit}$ results in the critical solution, super-critical initial conditions, which obtain for $p>p_\text{crit}$, exhibit scaling laws for the Schwarzschild radius and hence for the mass of the final black hole, of the form
\begin{equation}
r_S(p) \propto M_\text{bh}(p) \propto (p-p_\text{crit})^\gamma\,,
\end{equation}
where numerically one finds for the Choptuik exponent the value $\gamma\simeq 0.37$~\cite{Chop}. Similar observations were made in ~\cite{Hamade:1995ce,Gundlach:2002sx}.

Let us stress, however, that in a generic dimension $d \geq 4$ the definition must be generalized as follows~\cite{KHA,AlvarezGaume:2006dw}
\begin{equation}
 r_S(p) \propto (p-p_\text{crit})^\gamma \,, \quad M_\text{bh}(p) \sim (p-p_\text{crit})^{(D-3)\gamma}  \,.
\end{equation}
Various numerical experiments with different choices of matter content can be performed along the same lines. For instance~\cite{Birukou:2002kk,Husain:2002nk,Sorkin:2005vz,Bland:2005vu} found more types of critical solutions for a massless scalar field. The case of a complex scalar field was investigated in detail in ~\cite{HirschmannEardley}, and the critical collapse of a radiation fluid was considered in the interesting works~\cite{AlvarezGaume:2008qs,evanscoleman,KHA,MA}. For spherically symmetric radiation fluids~\cite{evanscoleman} obtained a very similar result, $\gamma \simeq 0.36$, which led to the natural conjecture \cite{Strominger:1993tt} that $\gamma$ ought to be universal for four-dimensional gravity coupled to any matter content.

It was then understood~\cite{KHA,MA,Hirschmann:1995qx} that the Choptuik exponent can be extracted more conveniently  relying on perturbations of the self-similar solutions. To this end, one perturbs a self-similar field $h_0$ according to
\begin{equation}
    h = h_0 + \varepsilon \, h_{-\kappa}
\end{equation}
where the perturbation $h_{-\kappa}$ has scaling dimension $-\kappa \in \mathbb{C}$, to then determine the values of $\kappa$ corresponding to the allowed modes. The most relevant mode $\kappa^*$, for which $Re(\kappa)$ attains the maximum value, is related to the Choptuik exponent according to
\begin{equation}
    \gamma = \frac{1}{\Re \kappa^*}\,.
\end{equation}
This approach allowed for more precise determinations of the Choptuik exponents. For the radiation fluid, one thus arrived~\cite{KHA} at a better estimate, $\gamma \simeq 0.3558$, which clearly differs from the value that was originally obtained for the real scalar field. The authors of~\cite{Hirschmann:1995qx} then determined the Choptuik exponent for a complex scalar field, obtaining the value $\gamma \simeq 0.3871$, which is even more difficult to reconcile with the original result for a real scalar field. Therefore, the original hope that the value of $\gamma$ reflect some sort of universality, independently of the matter content, was not fulfilled.

The extension to the case of axial symmetry was then considered in~\cite{AE}, while the authors of~\cite{AlvarezGaume:2008fx} examined critical collapse in the presence of gravitational shock waves. Non-linear sigma models provide an interesting class of slightly more complex systems. Specifically,~\cite{Hirschmann_1997} investigated critical collapse in four dimensions in the presence of sigma models related to the two-sphere and the hyperbolic plane $\mathbb{H}^2$, whose critical solutions are self-similar up to compensating $U(1)$ internal rotations. The $\mathbb{H}^2$ sigma model is particularly interesting, since it describes the coupling of an axion-dilaton system to gravity. The corresponding Choptuik exponent, $\gamma \simeq 0.2641$, is significantly lower.

This paper is devoted to the study of continuously self-similar solutions of the Einstein-axion-dilaton system in arbitrary dimensions. The internal $\sltr$ symmetry of the scalar sector has an interesting structure, due to the simultaneous presence of compact and non-compact one-parameter subgroups, and we thus generalize previous discussions considering all possible classes of continuously self-similar collapses. In this fashion, we discover several families of critical spacetimes in various dimensions.
In a previous work~\cite{ours} we have studied self-similar solutions to the Einstein-axion-dilaton system in several dimensions, and with the compensating internal transformation in the three conjugacy classes of $\sltra$, generalizing~\cite{AlvarezGaume:2011rk, hatefialvarez1307}. Surprisingly, we discovered that even for a given choice of dimension, matter content and conjugacy class of compensating transformation, entire families of physically-distinct self-similar solutions exist. For example, the axion-dilaton with $U(1)$-compensated dilations (i.e., ``elliptic'' class), which was first considered in~\cite{Hirschmann_1997} in four dimensions, exhibits \emph{three} distinct collapse solutions in five dimensions.

In this work we perturb the different classes of solutions discovered in~\cite{ours}, providing  accurate estimates for the Choptuik exponents. Our main conclusion is that the Choptuik exponents depend not only on the choices of matter Lagrangian and spacetime dimension, but they also \emph{attain different values for the different critical points that may arise within a given system}. 
We could reach these conclusions for the three conjugacy classes relying on a robust analytic and numerical formalism for linear perturbations of self-similar solutions in arbitrary dimensions, which we developed here and in~\cite{ours}. We are thus able to recover the known value~\cite{Hirschmann_1997} of $\gamma$ for the unique four-dimensional elliptic critical collapse,
\begin{equation}
    \gamma \sim 0.2641
\end{equation}
within numerical accuracy, which supports our confidence in the current setup.

\section{Self-similar Einstein-axion-dilaton solutions}

We begin by summarizing, in this section, the main results of~\cite{ours}. Our discussion starts with a description of the Einstein-axion-dilaton system, to then review the concept of continuously self-similar (CSS) configurations, the procedure for determining CSS solutions and the detailed solutions found in four and five dimensions in~\cite{ours}, which provide the backgrounds for the perturbations examined here.

\subsection{Einstein-axion-dilaton system and self-similar fields}\label{sec:unperturbed}

The Einstein-axion-dilaton system in $d$ spacetime dimensions ($d\geq 4$) consists of gravity coupled to an axion $a$ and a dilaton $\phi$, and is described by the action 
\begin{equation}
S = \int d^d x \sqrt{-g} \left( R - \frac{1}{2} \frac{ \partial_a \tau
\partial^a \bar{\tau}}{(\Im\tau)^2} \right) \; .
\label{eaction}
\end{equation}
which is the same as the low energy effective action of type II string theory \cite{Sen:1994fa,Schwarz:1994xn} where the axion-dilaton complex field $\tau \equiv a + i e^{-\phi}$. Note that the $\sltr$ symmetry
reduces to the $\mathrm{SL}(2,\mathbb{Z})$ subgroup once quantum effects are considered,
and this S-duality is conjectured to be a
non-perturbative symmetry of IIB strings~\cite{gsw,JOE,Font:1990gx}. From the action~\eqref{eaction} one obtains the equations of motion
\begin{equation}
\label{eq:efes}
R_{ab} = \tilde{T}_{ab} \equiv \frac{1}{4 (\Im\tau)^2} ( \partial_a \tau \partial_b
\bar{\tau} + \partial_a \bar{\tau} \partial_b \tau)\,,
\end{equation}
\begin{equation}\label{eq:taueom}
\nabla^a \nabla_a \tau +\frac{ i \nabla^a \tau \nabla_a \tau }{
\Im\tau} = 0 \,.
\end{equation}
In~\cite{ours} and in the present paper we will only consider spherically symmetric configurations and perturbations\footnote{We suspect it should be possible to prove that non-spherical perturbation modes (either with angular dependence, or with polarisation indices on the sphere) never include the most relevant modes. Intuitively, just as in time-translation invariant systems, the modes with angular momentum have increased energy, we expect that in a scale-invariant system angular momentum necessarily increases $-\Re\kappa$. If this is indeed the case, then for the sake of determining the critical exponent spherically-symmetric perturbations are sufficient.}. The general spherically symmetric metric in $d$ dimensions can be cast in the form
\begin{equation}
    ds^2 = (1+u(t,r)) (-b(t,r)^2 dt^2 + dr^2) + r^2 d\Omega_q^2 \,,
\end{equation}
\begin{equation}
    \tau = \tau(t,r)\,.
\end{equation}
where $q \equiv d-2$.

The assumption of continuous scale invariance for the metric translates into a homogeneous scaling of the line element under dilations, as follows. If
\begin{equation}
    (t,r)\rightarrow ( \Lambda t,\Lambda r)\,,\quad \Lambda>0
\end{equation}
then
\begin{equation}
    ds^2 \rightarrow \Lambda^2 ds^2\, ,
\end{equation}
which implies the metric functions are themselves scale-invariant, so that
\begin{equation}\label{eq:metricscaling}
    u(t,r) = u(z)\,,\quad b(t,r) = b(z)\,,
\end{equation}
where $z \equiv -r/t$ is a scale-invariant coordinate.
The scalar $\tau$, rather than being directly scale-invariant, can be invariant, more generally, up to an $\sltr$ transformation, so that
\begin{equation}\label{eq:tauscaling}
    \tau(t,r) \rightarrow M(\Lambda) \tau(t,r)\,.
\end{equation}
We term a configuration $(g,\tau)$ satisfying eqts.~\eqref{eq:metricscaling},~\eqref{eq:tauscaling} to be continuously self-similar (CSS).

Physically distinct cases correspond to the different conjugacy classes of $\eval{\dv{M}{\Lambda}}_{\Lambda=1}$. For the three cases of elliptic, parabolic and hyperbolic elements one can pa\-ra\-metrize the CSS axion-dilaton field as
\begin{equation}\label{eq:tauansatz}
    \tau(t,r) = \begin{dcases}
                   i \frac{1-(-t)^{i\omega}f(z)}{1+(-t)^{i\omega} f(z)}\,, & \quad \text{elliptic}\\[5pt]
                    f(z) + \omega \log(-t)\,, & \quad \text{parabolic}\\[5pt]
                    (-t)^\omega f(z)\,, & \quad \text{hyperbolic}
                \end{dcases}
\end{equation}
in terms of an unknown real parameter $\omega$ and a scale-invariant field $f(z)$, which satisfies $\abs{f(z)} < 1$ in the elliptic case and $\Im f(z)>0$ in the parabolic and hyperbolic cases. Note that $\omega$ can be assumed to be positive as the $\sltr$ symmetry $\tau\rightarrow - \frac{1}{\tau}$ leaves the form of the ans\"atze~\eqref{eq:tauansatz} invariant, but changes the sign of $\omega$.

At this point, making use of the CSS ans\"atze in the equations of motion~\eqref{eq:efes},~\eqref{eq:taueom}, one can recover the CSS equations $u(z)$, $b(z)$, $f(z)$. In fact, $u(z)$ can be eliminated in favour of $b(z)$ and $f(z)$, and after some simplifications one can construct a final system of ordinary differential equations (ODEs) 
\begin{align}\label{eq:unperturbedbp}
    b'(z) & = B(b(z),f(z),f'(z))\,, \\
    f''(z) & = F(b(z),f(z),f'(z))\,. \label{eq:unperturbedfpp}
\end{align}
These equations have several singular points; the relevant range for $z$, which contains the infinite past, lies between the two singularities
\begin{equation}
    z = 0\,,
\end{equation}
\begin{equation}
    z = z_+\,,\quad b(z_+) = z_+\,.
\end{equation}
In particular, $z=z_+$ is a null surface and is in fact a type of event horizon (homothetic horizon). This surface should be merely a coordinate singularity, and thus the axion-dilaton field $\tau$ should be regular across it. This condition is equivalent, in fact, to demanding that $f''(z)$ remain finite as $z\rightarrow z_+$; the vanishing of the divergent part of $f''(z)$ is a complex-valued constraint at $z_+$
\begin{equation}\label{eq:unpconstraint}
    \mathbb{C} \,\ni\, G(b(z_+), f(z_+), f'(z_+))\,.
\end{equation}
For what concerns $z=0$, using the regularity of $\tau$ and residual symmetries in the equations of motions, one can determine the initial conditions
\begin{equation}
        b(0) = 1\,,\quad f'(0) =0\,, 
\end{equation}        
\begin{equation}
        f(0) = \left\{\begin{array}{l l l}
        x_0 & \text{elliptic}       & 0<x_0<1 \\
        i x_0 & \text{parabolic} & 0<x_0\\
        1+i x_0 & \text{hyperbolic} & 0<x_0
    \end{array}\right.
\end{equation}
where $x_0$ is a real parameter. At this point, one is left with two unknown parameters $(\omega,x_0)$ and two constraints (the real and imaginary parts of $G$). The system is thus completely determined, and will generically feature a discrete solution set.
In~\cite{ours} we investigated this system numerically in four and five dimensions, finding several CSS solutions that we now review briefly.

\subsection{Solutions in four and five dimensions}\label{sec:bgsolutions}

In this section we review the CSS solutions in four and five dimensions for the three conjugacy classes of $\sltra$ found\footnote{No solutions were identified in the parabolic class, within our numerical precision.} in~\cite{ours}.  They are determined by the values of the parameters $\omega$ and $x_0$, as explained in Section~\ref{sec:unperturbed}, and we also note the position of the $z_+$ singularity. The actual profiles of these solutions can be readily reproduced integrating numerically the CSS equations of motion.

Notice that multiple distinct solutions were found in~\cite{ours} for some specific choices of dimensions and conjugacy classes. The set of $11$ CSS solutions in four and five dimensions discovered in~\cite{ours} corresponds to the following parameters:

\begin{minipage}{\textwidth}
\begin{center}
\begin{tabular}{|c|c|c|c|}
    \hline
     \multicolumn{4}{|c|}{\textbf{4d elliptic}}\\ \hline
      & $\omega$ & $\abs{f(0)}$ & $z_+$\\
      & $1.176$ &   $0.892$ & $2.605$\\\hline
\end{tabular}
\tablespacing
\begin{tabular}{|c|c|c|c|}
      \hline
      \multicolumn{4}{|c|}{\textbf{5d elliptic}}\\\hline
        & $\omega$ & $\abs{f(0)}$ & $z_+$\\
      $\alpha$ & $.999$ &   $0.673$ & $1.246$\\
      $\beta$& $1.680$ &   $0.644$ & $1.397$\\
      $\gamma$& $2.304$ &   $0.700$ & $1.694$\\\hline 
\end{tabular}
\tablespacing
\begin{tabular}{|c|c|c|c|}
    \hline
     \multicolumn{4}{|c|}{\textbf{4d hyperbolic}}\\ \hline
      & $\omega$ & $\Im f(0)$ & $z_+$\\
      $\alpha$ & $1.362$ & $0.708$ & $1.440$\\
     $\beta$ & $1.003$ & $0.0822$ & $3.29$ \\
     $\gamma$ & $0.541$ & $0.0059$ & $8.44$\\\hline
\end{tabular}
\tablespacing
\begin{tabular}{|c|c|c|c|}
\hline
      \multicolumn{4}{|c|}{\textbf{5d hyperbolic}}\\\hline
        & $\omega$ & $\Im f(0)$ & $z_+$\\
     $\alpha$ & $1.546$ & $1.555$ & $1.254$\\
    $\beta$ & $1.305$ & $3.086$ & $1.129$\\
    $\gamma$ & $1.125$ & $1.705$ & $1.109$\\
    $\delta$ & $0.588$ & $0.364$ & $1.156$\\\hline 
\end{tabular}
\end{center}
\end{minipage}

Let us also note that in~\cite{ours} we tentatively identified at least a fourth solution $\delta$ in the four-dimensional hyperbolic class. Since their numerical accuracy was very limited already in the unperturbed solutions, we exclude these cases from our current investigation. Moreover, we were unable to ascertain whether additional hyperbolic solutions were present in a region essentially inaccessible due to numerical errors. \cite{ours} contains a more precise description of the solution sets in parameter space, along with graphical representations.

\section{Perturbation theory}

In this section we would like to outline a comprehensive method  that can allow one to investigate perturbations in general numbers of dimensions and for any matter content. We apply this formalism to the axion-dilaton system and identify linearized perturbations of the equations of motions for elliptic and hyperbolic ans\"atze. We shall follow on the lines of previous discussion of the perturbation theory of self-similar backgrounds~\cite{Hamade:1995jx}\footnote{We also remark that~\cite{Ghodsi_2010} performs a similar analysis of perturbations of spherically symmetric background for a specific model.}. We will, however, improve upon the methodology by implementing several algebraic manipulations for the purpose of simplifying an otherwise extremely cumbersome and opaque computation.


We emphasize that by means of the following method we are able to reconstruct the well-known \cite{Eardley:1995ns} result for the Choptuik exponent for the four-dimensional elliptic solution, namely

\begin{equation}
    \gamma \sim 0.2641\,.
\end{equation}

We shall consider this as a strong indication that our algebraic approach, numerical procedure and implementation of the latter are all sound, and that additional results are to be trusted.

\subsection{Perturbation ans\"atze}

Given an exactly critical analytical solution $h$, one can perturb it to discover the critical exponent $\gamma$ letting
\begin{equation}
 h(t,r) = h_0(z) + \varepsilon \,h_1(t,r)
\end{equation}
where $\varepsilon$ is a small number. By expanding the equations of motion in powers of $\varepsilon$, the background equations are clearly the zeroth-order part, while the linearized equations for the perturbations $h_1(t,r)$ correspond to the linear terms in
$ \varepsilon $. The equations for the perturbations are by construction linear, and in addition they are themselves scale-invariant, which means they are autonomous in the $\log(-t)$ variable. This suggests the Fourier-Laplace decomposition
\begin{equation}
    h_1(t,r) = \sum_{\kappa} (-t)^{-\kappa} h_1^{(\kappa)}(z)\,,
\end{equation}
where $\kappa$ spans with generality over a discrete set of modes in $\mathbb{C}$. In $\kappa$-space, the equation one thus obtains is algebraic. We can therefore from now on omit the $(\kappa)$ index for $h_1^{(\kappa)}$, to consider directly perturbations of the form
\begin{eqnarray}\label{eq:generic_perturbation_ansatz}
h(z,t) = h_{0}(z) + \varepsilon (-t)^{- \kappa} h_{1}(z) \,,
\label{decay}
\end{eqnarray}
with the understanding that once one identifies the spectrum of $\kappa$ solving the corresponding equations for $h_1(z)$, the general solution to the first-order equations is given by a linear combination of these modes.

The spectrum of $\kappa$ will be generally not real. Modes with $\Re\kappa <0$ represent instabilities, and their existence would imply that the solution is not truly critical. Unstable CSS collapses were identified for example in~\cite{Hirschmann_1997} in four-dimensional $\sigma$-models. Given that the axion-dilaton system in four dimensions in the elliptic class is known to be stable, we have assumed that the same property holds for all cases studied in this work, and we have only confined our attention to $\Re \kappa >0$.

We are actually interested specifically in the mode $\kappa^*$ with largest real part, or in other words the fastest decaying mode, which is related to the Choptuik exponent via~\cite{KHA,MA,Hirschmann:1995qx}
\begin{eqnarray}\label{eq:chopkappa}
\gamma = \frac{1}{\Re \kappa^*}
\end{eqnarray}

We have assumed that, as for the four-dimensional elliptic critical collapse, $\kappa^*$ is real. This restriction is required due to the demanding computational workload, although some of the self-similar backgrounds discussed here might have complex fastest decaying modes. If so, our determination of the Choptuik exponent would be an overestimate.

\subsubsection{Pure gauge modes}

It is important to identify which values of $\kappa$ correspond to unphysical gauge modes. These will be solutions of the perturbed equations that however are not physically distinguishable from the background. One can generate these modes, and thus identify them, performing infinitesimal gauge transformations on the self-similar solutions. We remark that the following discussion is valid in any dimension.

Consider first the symmetry transformations $f(z) \rightarrow e^{i\epsilon} f(z)$ for the elliptic class, and $f(z)\rightarrow e^{\epsilon} f(z)$ for the hyperbolic class. If $\epsilon \ll 1$, they reduce to
\begin{equation}
    f_0(z) \rightarrow f_0(z) + \begin{cases}
            i \epsilon f_0(z) & \text{elliptic}\\
            \epsilon f_0(z) & \text{hyperbolic}
            \end{cases}
\end{equation}
while $b_0(z)$ is unaffected. Comparing with the perturbation ans\"atze~\eqref{eq:generic_perturbation_ansatz}, one can deduce that the following mode has been generated:
\begin{eqaed}
    (-t)^{-\kappa} f_1(z) & = \begin{cases}
            i  f_0(z) & \text{elliptic}\\
             f_0(z) & \text{hyperbolic}
            \end{cases}\,\\
    (-t)^{-\kappa} b_1(z) & = 0\,.
\end{eqaed}
One can thus see that this unphysical mode corresponds to $\kappa = 0$, and consequently this root must be excluded from the analysis.

In fact, a similar mode would also be present in the parabolic class, if one were to identify self-similar solutions there, but as we have anticipated we have found no solutions belonging to this class. At any rate, under $f(z)\rightarrow f(z)+\epsilon$ one would generate the mode $b_1(z)=0$, $f_1(z) = 1$, which has also $\kappa = 0$. Even more generally, in any model with a matter sector possessing internal symmetries, these will always give rise to pure gauge perturbations that are scale-invariant, and thus again with $\kappa=0$.

Time translations at constant $r$, $t \rightarrow t+ \epsilon$, are another class of gauge transformations. In this case $r\rightarrow r$, and therefore $z \rightarrow z - \frac{z}{t}\epsilon$, so that
\begin{eqaed}
    b_0(z) & \rightarrow b_0(z) - \frac{z}{t} \epsilon b_0'(z)\,,\\
    f_0(z) & \rightarrow f_0(z) - \frac{z}{t} \epsilon f_0'(z)\,.
\end{eqaed}
Comparing with the ansatz, one can read off again the value $\kappa = 1$. Therefore, this represents an additional unphysical gauge mode.

\subsection{Linearized equations of motion in any dimension}

We shall now set up the perturbation theory for the self-similar solutions discussed above. We first perform this step for the elliptic class of solutions in an arbitrary dimension $d = q+2 \geq 4$ 

\subsubsection{Linearized equations for the elliptic class}

The perturbation ansatz~\eqref{eq:generic_perturbation_ansatz}, specified to the $u$, $b$, $\tau$ functions, reads
\begin{equation}\label{eq:pertu}
    u(t,r) = u_0(z) + \varepsilon \, (-t)^{-\kappa} u_1(z) \ ,
\end{equation}
\begin{equation}\label{eq:pertb}
    b(t,r) = b_0(z) + \varepsilon\, (-t)^{-\kappa} b_1(z)\ ,
\end{equation}
\begin{equation}\label{eq:pertelliptictau}
    \tau(t,r) = i\frac{1-(-t)^{i\omega}f(t,r)}{1+(-t)^{iw}f(t,r)}\ ,
\end{equation}
\begin{equation}\label{eq:pertf}
    f(t,r) \equiv f_0(z) + \varepsilon (-t)^{-\kappa} f_1(z)\ .
\end{equation}
Using the ans\"atze~\eqref{eq:pertu},~\eqref{eq:pertb} for the metric functions, one can compute the Ricci tensor for the metric
\begin{equation}
 ds^2 = (1+u)( -b^2 dt^2 + dr^2) + r^2 d\Omega_q^2   
\end{equation}
as a function of $\varepsilon$. The dependence on the spacetime dimension $d = q+2$ can be determined following the procedure described in~\cite{ours}. One can thus find the zeroth-order and first-order parts from the limiting behaviors
\begin{equation}
    R^{(0)}_{ab} = \lim_{\varepsilon\rightarrow 0} R_{ab}(\varepsilon)\,,
\end{equation}
\begin{equation}
    R^{(1)}_{ab} = \lim_{\varepsilon\rightarrow 0} \dv{R_{ab}(\varepsilon)}{\varepsilon}\,.
\end{equation}
The same can be done with the right-hand side $\tilde{T}_{ab}$ of the field equations, introducing the axion-dilaton perturbations~\eqref{eq:pertelliptictau},~\eqref{eq:pertf}, and therefore, in a similar fashion,
\begin{equation}
    \tilde{T}^{(0)}_{ab} = \lim_{\varepsilon\rightarrow 0} \tilde{T}_{ab}(\varepsilon)\,,
\end{equation}
\begin{equation}
    \tilde{T}^{(1)}_{ab} = \lim_{\varepsilon\rightarrow 0} \dv{\tilde{T}_{ab}(\varepsilon)}{\varepsilon}\,.
\end{equation}
The Einstein Field Equations (EFEs) must hold order by order, so that
\begin{equation}
    R^{(0)}_{ab} = \tilde{T}^{(0)}_{ab}\,,\quad R^{(1)}_{ab} = \tilde{T}^{(1)}_{ab} \ .
\end{equation}
It is now possible to use part of these equations to eliminate $u(t,r)$ from the others. Specifically, one can use $R^{(0)}_{tr} = \tilde{T}^{(0)}_{tr}$ to eliminate $u_0'(z)$, $R^{(0)}_{ij} = \tilde{T}^{(0)}_{ij} = 0$ to eliminate $u_0(z)$, $R^{(1)}_{tr} = \tilde{T}^{(1)}_{tr}$ to remove $u_1'(z)$, and finally $R^{(1)}_{ij} = \tilde{T}^{(1)}_{ij} = 0$ to eliminate $u_1(z)$. 
For the sake of simplicity, from now on we shall not display the $z$ argument of all functions. In detail
\begin{align}
      \frac{q u_0'}{2(1+u_0)} &=  \frac{4zf_0'\bar f_0'+2i( \omega \bar f_0 f_0' - \omega f_0\bar f_0')}{2(1-f_0\bar f_0)^2} \, ,\\
    u_0 &= \frac{zb_0'}{(q-1)b_0} \,, \\
    u_1& =-\frac{(q-1)b_1u_0-zb_1'}{(q-1)b_0}\,.
\end{align}
We omit the explicit form of $u_1'(z)$, since it is rather cumbersome, and henceforth we shall also assume that $u(t,r)$ and its first derivatives have been expressed, in all equations, in terms of the other variables.
One is thus left with $b(t,r)$ as the only remaining metric function, and it is now interesting to combine the field equations in such a way as to eliminate the second-derivative terms in $b(t,r)$. This procedure identifies the so-called Hamiltonian constraint, which here results from the combination
\begin{equation}
    C(\varepsilon) \equiv R_{tt} + b^2\,R_{rr} - \tilde{T}_{tt} - b^2\, \tilde{T}_{rr} = 0\,.
\end{equation}
The lowest-order contribution is a first-order equation linking $b_0'$ to $b_0$ $f_0$, $f_0'$ (in fact, one thus recovers~\eqref{eq:unperturbedbp}),
\begin{eqnarray}\label{eq:ellipticb0p}
b_0'= \frac{2((z^2-b_0^2)f_0'(z \bar f_0'+i \omega  \bar f_0)+\omega f_0(\omega z \bar f_0-i(z^2-b_0^2) \bar f_0'))}{q b_0 (1-f_0 \bar f_0)^2}\,.
\end{eqnarray}
Note that the right-hand side of~\eqref{eq:ellipticb0p} is real, as expected. In a similar fashion the first correction,

\begin{equation}
    \eval{\dv{C(\varepsilon)}{\varepsilon}}_{\varepsilon=0} =0 \ ,
\end{equation}

is a first-order equation linking $b_1'$ to $b_0$, $b_0'$ $f_0$, $f_0'$, $f_0''$, and the other perturbations $b_1$, $f_1$, $f_1'$, which is of course linear in all perturbations. For the elliptic class this reads
\begin{eqaed}\label{eq:equationb1p_elliptic}
(L_1)b_1'&=r t ((t-q t)b_0+rb_0')\bigg(-\frac{2 f_1' \bar f_0' r^2}{t^4 s_0^2}-\frac{q b_1 b_0'}{r t}+\frac{\kappa q b_0 b_1 b_0'}{r \left((t-q t) b_0+rb_0'\right)}\\& \quad
-\frac{4 i \omega b_0 b_1 \bar f_0 f_0'}{r t s_0^2}
+\frac{2i \omega b_0^2 \bar f_1 f_0'}{r t s_0^3}
+\frac{2\kappa \bar f_1 f_0'r}{t^3 s_0^2}+\frac{2i  \omega \bar f_1 f_0' r}{t^3 s_0^2}
\\& \quad 
-\frac{2 \kappa b_0^2 \bar f_1 f_0'}{r t s_0^2}
+\frac{2i \omega \bar f_0  f_1'r}{t^3 s_0^2}
-\frac{2 i \omega b_0^2 \bar f_0 f_1'}{r t s_0^2}
+\frac{4 b_0 b_1 f_0' \bar f_0'}{t^2 s_0^2}
 \\& \quad 
+\frac{2b_0^2 f_1' \bar f_0'}{t^2 s_0^2}
+\frac{1}{r t^4 s_0^3} 2f_1\Big(t \omega \bar f_0^2 (rt(-i\kappa+\omega)f_0-2i(r^2-t^2b_0^2)f_0')
\\& \quad \quad
+t(\kappa-i\omega)(-r^2+t^2b_0^2) \bar f_0'
+\bar f_0 \big(rt^2 \omega(i\kappa+\omega)+t(\kappa+i\omega)
\\& \quad \quad \times(r^2-t^2b_0^2)f_0\bar f_0'
+2r (r^2-t^2b_0^2)f_0'\bar f_0'\big)\Big)
-\frac{2 f_0' \bar f_1' r^2}{t^4 s_0^2}
+\frac{2 b_0^2 f_0' \bar f_1'}{t^2s_0^2} \\
& \quad -\frac{2if_0 (\bar f_1(r\omega t^2 (\kappa+i\omega)+\omega t(2r^2-t^2 b_0^2)\bar f_0 f_0'+2ir(r^2-t^2 b_0^2) f_0' \bar f_0'))}{rt^4s_0^3}
\\& \quad
-\frac{2 i f_0 t \omega(-\bar f_1' r^2+2t^2 b_0 b_1 \bar f_0'+t^2 b_0^2 \bar f_1')}{rt^4s_0^3}
+\frac{4 i \omega f_0^2 \left(r^2-t^2 b_0^2\right) \bar f_1 \bar f_0'}{r t^3 s_0^3}
\\&
+\frac{2 i \omega f_0^2 \left(\bar f_0 \left(-\bar f_1' r^2+t (\kappa-i \omega) \bar f_1 r+2 t^2 b_0 b_1 \bar f_0'+t^2b_0^2 \bar f_1'\right)\right)}{r t^3 s_0^3}\bigg)\,,
\end{eqaed}
where
\begin{equation}
    L_1 = q b_0 ((\kappa-q+1) tb_0+rb_0'), \quad  s_0 =(f_0 \bar{f}_0-1) \ .
\end{equation}
At this point we can make an important remark: eq.~\eqref{eq:equationb1p_elliptic} is evidently invariant under dilations $(t,r)\rightarrow (e^\lambda r,e^\lambda t)$, so that, changing coordinates to $(t,z)$, the factors of $t$ will cancel and the result will only depends on $z$. All in all, one is actually dealing with a \emph{real} and \emph{linear ordinary differential equation}. 

One can now introduce the perturbation ans\"atze in the $\tau$ equation of motion~\eqref{eq:taueom}. After extracting the zeroth and first-order parts, one can finally replace all instances of $u_0$, $u_0'$, $u_1$, $u_1'$ as above. The resulting zeroth-order part is an equation involving $b_0$, $b_0'$ $f_0$, $f_0'$, $f_0''$, while replacing $b_0'$ according to eq.~\eqref{eq:unperturbedbp} and solving for $f_0''$, one finally arrives at a second-order equation for $f_0$, which is the explicit form of eq.~\eqref{eq:unperturbedfpp}:
\begin{eqaed}
L_2 f_0''&=z^2 b_0' s_0(z f_0'-i \omega f_0)+z b_0^2 b_0' s_0 f_0'
\\
&-b_0^3 f_0'(2 z \bar f_0 f_0'-q f_0 \bar f_0+q)
 \\& -zb_0\Big(f_0\big(2(1+i \omega)z \bar f_0f_0'+\omega (\omega+i)\big)
 \\& \quad -2zf_0'(z\bar f_0f_0'-i \omega+1)+\omega(\omega-i)f_0^2 \bar f_0\Big)\,. \end{eqaed}
Here
\begin{equation}L_ 2= zb_0(z^2-b_0^2) s_0\,,\end{equation}
and again the first-order part will involve $b_1$, $b_1'$, $f_1$, $f_1'$, $f_1''$ linearly, and the zeroth-order functions and their derivatives non-linearly. In detail
\begin{eqaed}
  (L_3)f_1''=&r^2\bigg( it \omega f_0^2(\bar f_1b_0'+\bar f_0 b_1') +f_1b_0'(\kappa t-it \omega -t(\kappa-2i\omega)f_0\bar f_0+r f_0'\bar f_0)
  \\
  &\quad -r( b_1'f_0'+b_0'f_1')+f_0\big(b_1'(-it \omega+r\bar f_0 f_0')+rb_0'(\bar f_1 f_0'+\bar f_0 f_1')\big)\bigg)\\
  & +rt^2b_0^2\big(f_1\bar f_0 b_0'  f_0'-b_1' f_0'-b_0' f_1'+f_0(\bar f_1 b_0' f_0'+\bar f_0(b_1'f_0'+b_0'f_1'))\big)\\
  & -t^2b_0^3\bigg(2r\bar f_1f_0'^2-qtf_1'+4r\bar f_0 f_0' f_1'+f_1 \bar f_0(qtf_0'-rf_0'')\\
  & \quad  +f_0\big(qt\bar f_0 f_1'+\bar f_1(qtf_0'-rf_0'')\big)\bigg)\\
  &-rb_0\bigg(t^2\omega(\omega-i)f_0^2\bar f_1 +2r\big(-r \bar f_1 f_0'^2+(t(1+\kappa-i\omega)-2r\bar f_0 f_0')f_1'\big)\\
  & \quad \quad  +f_1\Big(t^2(-\kappa^2+\kappa(-1+2i \omega)+ \omega (i+\omega))+t^2(\kappa+\kappa^2+2i\kappa\omega \\
  &  \quad \quad \quad +2\omega(-i+\omega)\big)\bar f_0 f_0+r\bar f_0(2t(-1+2\kappa-i\omega)f_0'+rf_0'')\Big)\\
  & \quad \quad +rf_0(-2t(1+\kappa+i\omega)\bar f_0 f_1'+\bar f_1(-2it(-i+\omega)f_0'+rf_0''))\bigg) \\
  & -b_1\bigg(rt^2 \omega(-i+i\kappa+\omega)f_0^2\bar f_0-t\big(r^2(-2+\kappa+2i\omega)+3qt^2b_0^2\\
  &\quad -2rtb_0b_0'\big)f_0'
  -2r(r^2-3t^2b_0^2)\bar f_0f_0'^2-r(r^2-3t^2b_0^2)f_0''
  \\
  &\quad +f_0\Big(rt^2\omega(i-i\kappa+\omega)+
  \bar f_0\big(t(r^2(-2+\kappa-2i\omega)+3qt^2b_0^2\\
  &\quad -2rtb_0b_0')f_0'
  +r(r^2-3t^2b_0^2)f_0''\big)\Big)\bigg)\,,
  \end{eqaed}
where
\begin{equation}
    L_3= r b_0(r^2-t^2b_0^2) s_0\,.
\end{equation}
Note, again, that the equation is explicitly scale-invariant, and therefore it turns into an ordinary differential equation after a change of coordinates to $(z,t)$.

Once the zeroth-order solutions are obtained integrating numerically the unperturbed equations, and after replacing $b_1'$ via eq.~\eqref{eq:equationb1p_elliptic}, one is left with a system of ordinary linear differential equations that we write schematically in the form
\begin{align}
    b_1' & = B_1(b_1,f_1,f_1')\,,\label{eq:equationb1p}\\
    f_1'' & = F_1(b_1,f_1,f_1')\,.\label{eq:equationf1pp}
\end{align}
$B_1$ and $F_1$ are linear functions that still depend non-linearly on the unperturbed solution. In addition, there is still a quadratic dependence on $\kappa$. Note that these equations inherit the singularities of the unperturbed system of eqs.~\eqref{eq:unperturbedbp} and~\eqref{eq:unperturbedfpp}. In particular, they are also singular for $z=0$ and $b^2(z)=z^2$.

\subsubsection{ Linearized equations for the Hyperbolic class}

The same procedure can be followed for the hyperbolic class, again in an arbitrary number of dimensions. In this case eq.~\eqref{eq:pertelliptictau} has to be replaced by
\begin{equation}
    \tau(t,r) = (-t)^\omega f(t,r)\, ,
\end{equation}
while all the other assumptions are unchanged. The form of $u_0$ and $u_1$ is the same as in the elliptic case, since these quantities are independent of the matter content. One thus obtains for the  hyperbolic class of solutions the following form for $u_0'$:
\begin{eqnarray}
\frac{q u_0'}{2(1+u_0)}=  \frac{w(f_0 \bar f_0' + f_0' \bar f_0) - 2 z f_0' \bar f_0'}{(f_0 - \bar f_0)^2} \,.\end{eqnarray}
%
%
As before, for brevity we omit the explicit expression of $u_1'(z)$.

Following similar steps as before, the unperturbed equations of motion in the hyperbolic class are found to be
\begin{equation}
b_0'= -\frac{2 \left(\left(z^2-b_0^2\right) f_0' \left(z \bar f_0'-\omega \bar f_0\right)+\omega f_0 \left(\left(b_0^2-z^2\right) \bar f_0'+\omega z \bar f_0\right)\right)}{q b_0 (f_0-\bar f_0)^2}\,,
\end{equation}
\begin{eqaed}
L_4 f_0''& = z b_0^2(f_0 -\bar f_0) b_0' f_0'+b_0^3 f_0'
(qf_0 -q\bar f_0-2z f_0') \\& + z^2(f_0 -\bar f_0) b_0'  (-\omega f_0+zf_0') \\&  + z b_0\big( \omega (1+\omega)f_0^2+2z f_0'(\bar f_0-\omega \bar f_0+zf_0') \\& \quad  + f_0((\omega-1) \omega \bar f_0-2(1+\omega)zf_0')\big)\,,
\end{eqaed}
where  $L_4= zb_0(z^2-b_0^2) (f_0- \bar f_0)$. 

Retracing the steps of the previous section yields the linearized first-order equation for $b_1'$
\begin{eqaed}
(L_5)b_1'&= r((t-qt)b_0+rb_0')\Big(-m_0^2 2t^2(\kappa -\omega) \omega f_1\bar f_0 +m_0^2 2t^2 \omega(-\kappa +\omega)f_0\bar f_1\\& -\frac{qt^3b_1b_0'}{r} + \frac{\kappa qt^4b_0b_1b_0'}{r((t-qt)b_0+rb_0')}
+m_0^2 2rt(-\kappa +\omega)\bar f_1 f_0'+\frac{m_0^2 2t^3(\kappa -\omega)b_0^2\bar f_1 f_0'}{r}
\\
&+ m_0^2 2rt\omega \bar f_0 f_1'
- m_0^2 2rt(\kappa -\omega)\bar f_0' f_1+
\frac{m_0^2 2t^3(\kappa -\omega)b_0^2\bar f_0' f_1}{r}
+m_0^2 4t^2 b_0 b_1\bar f_0' f_0' \\
& +m_0^3 4t^2 b_0^2(\bar f_1-f_1)\bar f_0' f_0'
+\frac{m_0^3}{r}(4t^2b_0^2(f_1-\bar f_1)(t \omega f_0+rf_0')\bar f_0')+m_0^2 2r^2\bar f_0' f_1'\\
&+\frac{1}{r}m_0^3( 4t^2b_0^2(f_1-\bar f_1)f_0'(t\omega \bar f_0+r\bar f_0'))-m_0^3(4(f_1-\bar f_1)(t\omega f_0+rf_0')(t\omega \bar f_0+r\bar f_0'))
\\&-\frac{1}{r}m_0^2 2t^2b_0^2f_1'(t\omega \bar f_0+r\bar f_0')
-\frac{1}{r}m_0^2(4t^2b_0b_1(t\omega \bar f_0 f_0'+(t\omega f_0+2rf_0') \bar f_0'))
\\&
+m_0^2 2rt\omega f_0 \bar f_1'+m_0^2 2r^2 f_0' \bar f_1'-\frac{1}{r}m_0^2 2t^2b_0^2(t\omega f_0+rf_0')\bar f_1'\Big)\,,
\end{eqaed}
where $L_5= q t^3 b_0((1+\kappa -q)t b_0+rb_0')$ and $m_0=\frac{1}{(f_0-\bar f_0)}$.

Finally, the linearized equation for $f_1''$ reads
\begin{eqaed}
(L_6)f_1''&= \bigg(t^2( \kappa ^2-2\kappa \omega+\kappa +(\omega-1)\omega)b_0f_1-2t^2b_0b_1b_0'f_0'-t^2b_0^2b_1'f_0'
 \\
 & +(\kappa tb_1-rb_1')(t\omega f_0+rf_0')
 -(2b_1(t^2 \omega^2f_0^2+2rt\omega f_0f_0'+(r^2-t^2b_0^2)f_0'^2))m_0
 \\
 & -\big(2b_0(f_1-\bar f_1)(-t^2 \omega^2f_0^2-2rt\omega f_0f_0'+(-r^2+t^2b_0^2)f_0'^2)\big)m_0^2
 \\
 & -2rt(1+r-\omega)b_0f_1'+\frac{qt^3b_0^3f_1'}{r}-t^2b_0^2b_0'f_1'-rb_0'(t(-\kappa +\omega)f_1+rf_1')
 \\
 & +m_0 4b_0\big(-t\omega f_0(t(-\kappa +\omega)f_1+rf_1')+f_0'(rt(\kappa -\omega)f_1
 \\
 & +t^2b_0b_1f_0'-r^2f_1'+t^2b_0^2f_1')\big)
 +\frac{3t^2b_0^2b_1(qtf_0'-rf_0'')}{r}
 \\
 & +b_1(t^2(\omega-1)\omega f_0+r(2t(\omega-1)f_0'+rf_0''))\bigg)\,,
\end{eqaed}
where $L_6= -r^2 b_0+t^2b_0^3 $.
Once more, the equations for $b_1$ and $f_1$ are scale-invariant, and thus can be turned into ordinary differential equations in $z$.
In both cases, the modes are identified determining the values of $\kappa$ that lead to smooth solutions of the perturbed equations~\eqref{eq:equationb1p},~\eqref{eq:equationf1pp} that satisfy certain boundary conditions, which we now turn to describe.

\subsection{Boundary conditions for perturbations}

Eqs.~\eqref{eq:equationb1p},~\eqref{eq:equationf1pp} are to be supplemented by suitable boundary conditions. At $z=0$, thanks to the freedom to rescale the time coordinate, one can set
\begin{equation}
    b_1(0) = 0\,.
\end{equation}
In addition, regularity of the axion-dilaton implies
\begin{equation}
    f_1'(0) = 0\,.
\end{equation}
So that one is left only with an unknown complex parameter, $f_1(0)$.

At $z_+$ one can also demand that the axion-dilaton (and, thus, its perturbation) be regular across the homothetic horizon, which we emphasize is only a coordinate singularity. We argue that regularity is encoded in the finiteness of the second derivatives $\partial_r^2 f(t,r)$, $\partial_r \partial_t f(t,r)$, $\partial_t^2 f(t,r)$ as $z\rightarrow z_+$, since $t$ and $r$ are themselves regular there. This implies that $f_0''(z)$ and $f_1''(z)$ must be finite as $z \rightarrow z_+$. In practice, we define a parameter $\beta = b_0(z)-z$ and then expand $f_0''$ and $f_1''$ near the singularity, as
\begin{align}\label{eq:taylorexpansion_unperturbed}
    f_0''(\beta) & = \frac{1}{\beta} G(h_0) + \mathcal{O}(1)\,,\\ \label{eq:taylorexpansion_linearised}
    f_1''(\beta) & = \frac{1}{\beta^2} \bar{G}(h_0) + \frac{1}{\beta} H(h_0, h_1|\kappa) + \mathcal{O}(1)\,,
\end{align}
where we have introduced the schematic notation $h_0 = (b_0(z_+),f_0(z_+),f_0'(z_+))$, $h_1 = (b_1(z_+),f_1(z_+),f_1'(z_+))$. The vanishing of $G(h_0)$ is simply the unperturbed complex value constraint~\eqref{eq:unpconstraint} at $z_+$, and we shall assume that background solutions satisfy this condition. We also verify that
\begin{equation}
    G(h_0) = 0 \,\Rightarrow \, \bar{G}(h_0) = 0\,,
\end{equation}
which is a consistency check. One is thus only left with the complex-valued constraint
\begin{equation}\label{eq:constraintH}
    H(h_0, h_1 | \kappa) = 0\,,
\end{equation}
which is linear in $h_1$. We do not display the specific forms of this constraint for the elliptic and hyperbolic classes, since they are particularly unwieldy, but we remark that they are in principle easily obtained from the zeroth-order and first-order equations of motion according to eqs.~\eqref{eq:taylorexpansion_unperturbed} and~\eqref{eq:taylorexpansion_linearised}. We solve this constraint for $f_1'(z_+)$ in terms of $f_1(z_+)$, $b_1(z_+)$, $\kappa$ and $h_0$. This condition has thus reduced the number of free parameters in the boundary conditions at $z_+$ to a real number $b_1(z_+)$ and a complex $f_1(z_+)$.
Summarizing, there are six real unknowns, which are $\kappa$ and the five-component vector:
\begin{equation}\label{eq:defX}
    X = (\Re f_1(0),\, \Im f_1(0),\, \Re f_1(z_+),\, \Im f_1(z_+),\, b_1(z_+) )
\end{equation}
and the system of linear ODE's eqs.~\eqref{eq:equationb1p},~\eqref{eq:equationf1pp}, whose total real order is five.

\subsection{Numerical procedure}

We define a ``difference function'' $D(\kappa;X)$ ($X$ is as defined in~\eqref{eq:defX}) through the following procedure:

\begin{itemize}
    \item First solve the linear equation~\eqref{eq:constraintH} to determine $f_1'(z_+)$.
    \item With complete boundary conditions at $z=0$ and $z=z_+$, integrate forward from $z=0$ to an intermediate point $z_\text{mid}$, and also backward from $z_+$ to $z_\text{mid}$.
    \item Collect in a vector $(b_1, \Re f_1, \Im f_1, \Re f_1', \Im f_1')$ computed at $z_\text{mid}$ the difference between the two results.
\end{itemize}

We note that $D(\kappa; X)$ is linear in $X$. Indeed, since~\eqref{eq:constraintH} is linear in $X$, so are the boundary conditions for the ODE system, and since the solutions to a linear ODE system are linear in the boundary conditions themselves, so will be the differences of the values at $z_\text{mid}$ as functions of $X$. Since $D$ is linear, it admits a representation as a matrix product:
\begin{equation}
    D(\kappa; X) = A(\kappa) X
\end{equation}
with $A(\kappa)$ a $5\times 5$ real matrix depending non-linearly on $\kappa$.

The search for solutions is thus reduced to the task of finding zeroes of $D(\kappa; X)$. Note that, for any value of $\kappa$, non-trivial solutions in $X$ to the linear equation
\begin{equation}\label{eq:Dzero}
    D(\kappa; X) = A(\kappa) X = 0
\end{equation}
exist if and only if
\begin{equation}\label{eq:detzero}
    \det A(\kappa) = 0\,.
\end{equation}
Eq.~\eqref{eq:detzero} is a real equation for one real parameter. It is all what is needed to identify modes, and it is not necessary to find the actual form of the solutions.

We compute the columns of $A(\kappa)$ numerically through evaluations on the basis vectors of $\mathbb{R}^5$:
\begin{equation}
    A_j = D(\kappa;e_j)\,,
\end{equation}
which amounts to five evaluations of $D(\kappa;X)$. We then perform an easy root search for the determinant as a function of $\kappa$. The root with the largest value is related to the Choptuik exponent through eq.~\eqref{eq:chopkappa}.

\subsubsection{Momentum-space singularity}

One could notice that the perturbed equations of motion~\eqref{eq:equationb1p_elliptic}, \eqref{eq:equationf1pp} are singular whenever the factor
\begin{equation}
    W = \left(\kappa+1-q -z \frac{b_0'}{b_0}\right)\,,
\end{equation}
present in the denominators, vanishes.
For fixed $\kappa$, a zero can be encountered in the integration, and the numerical procedure fails at that point. We actually find that there are entire regions for $\kappa$, in the different solutions, which lead to failure. We can roughly pinpoint the location of these intervals determining the value of $\kappa$ for which a singularity appears directly at $z_+$:
\begin{equation}
    \eval{W}_{z = z_+} = \kappa+1-q- b_0'(z_+)\,,\quad \kbad = q-1 +b_0'(z_+)\,.
\end{equation}
In these case the divergence affects directly the boundary condition~\eqref{eq:constraintH}.

We find evidence that there is generically a whole range of values for $\kappa$ near $\kbad$, within which the singularity is reached somewhere between $z=0$ and $z=z_+$. $\kbad$ appears to usually lie either within or at the right edge of this interval. The numerical procedure more specifically either fails as the integration cannot be brought to completion or because numerical instabilities are not properly taken into account, and the output of the difference function is then incorrect.

One could possibly bypass this issue performing in advance suitable algebraic reparametrizations. However, the most relevant mode $\kappa^*$ lies, in almost all cases, outside the failure region (either before or after), so that this problem does not affect our estimates of the Choptuik exponents (see~Table~\ref{tab:results}).

\section{Results}

We can now present the numerical results for the Choptuik exponents of the critical solutions, in four and five dimensions, in the elliptic and hyperbolic classes introduced in Section~\ref{sec:bgsolutions}.

\begin{figure}
    \centering
    \includegraphics[width=4in]{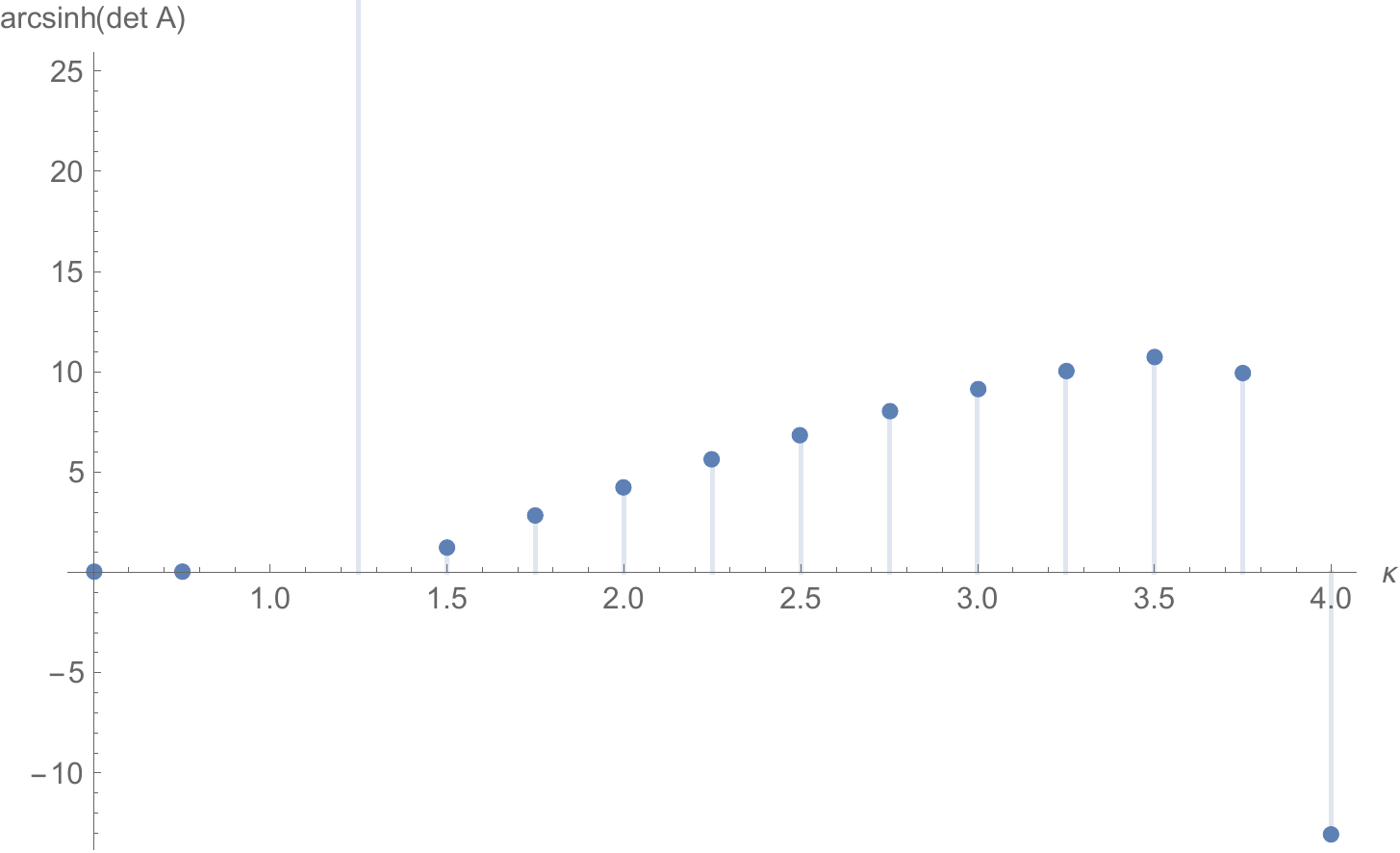}
    \caption{Behaviour of the determinant~\eqref{eq:Dzero} as a function of $\kappa$ for the four-dimensional elliptic solution. For clarity, we actually plot $\operatorname{arcsinh}(\det A(\kappa))$ in order to limit the range of values.}
    \label{fig:choptuik_large}
\end{figure}

\begin{figure}
    \centering
    \includegraphics[width=4in]{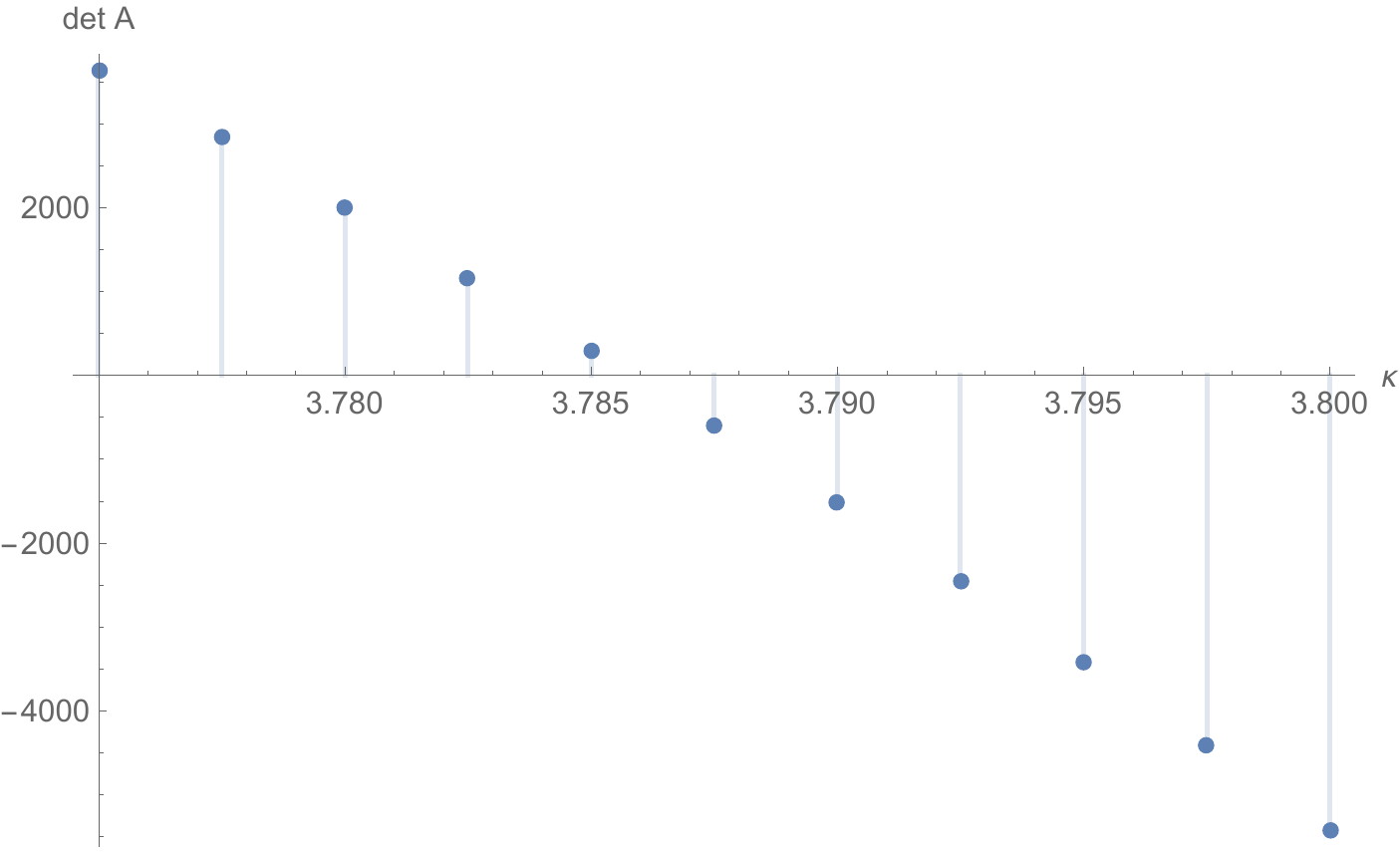}
    \caption{Behaviour of $\det A$ near the last crossing in the elliptic four-dimensional case.}
    \label{fig:choptuik_small}
\end{figure}

We first test the technique on the unique four-dimensional elliptic solution (4E). In Figs.~\ref{fig:choptuik_large}, \ref{fig:choptuik_small} we display the behaviour of $\det A(\kappa)$ near the last crossing of the horizontal axis. The position of the crossing is estimated with some precision to be
\begin{equation}
    \kappa^*_{4E} \approx 3.7858\,,
\end{equation}
which implies a Choptuik exponent
\begin{equation}
    \gamma_{4E} \approx 0.2641\,,
\end{equation}
although, strictly speaking, our determination is guaranteed to be precise only up and including the first two decimal digits\footnote{We base our estimate of the uncertainty in the Choptuik exponents on our ability to reconstruct the actual position of the $\kappa = 1$ gauge mode.}.

Remarkably, this result is in perfect agreement with the existing literature, see \cite{Eardley:1995ns}. We take this non-trivial match as evidence that our methodology and our numerical set up are reliable.
For this solution, $\kbad = 1.224$, and the integration fails around $1 \lesssim \kappa \lesssim 1.4$, well below the position of the most relevant mode. This is visible as a range of $\kappa$ values in Fig.~\ref{fig:choptuik_large} for which Mathematica could not complete the computation of $\det A$ or returned a grossly incorrect output. In any case, because of the considerable distance of this region from the final crossing, we have reasons to consider our numerical determination of $\kappa^*$ reliable.

Our results for the other Choptuik exponents can be found in Table~\ref{tab:results}, together with some additional details on the failure regions.

Note that our results are in contrast with those of~\cite{AlvarezGaume:2011rk, hatefialvarez1307}. In fact, as there exists a mismatch already at the level of the self-similar solutions between \cite{ours} and~\cite{AlvarezGaume:2011rk, hatefialvarez1307}, there is no expectation of overlap of the results of perturbation theory. 
\begin{center}
\begin{table}[ht]
\begingroup
\renewcommand*{\arraystretch}{2}
\begin{tabular}{|c|c|c|c|c|c|}
\hline
solution & $\kappa^*$ & $\gamma$ &\begingroup \renewcommand*{\arraystretch}{1}\begin{tabular}{@{}c@{}}failure\\region\end{tabular} \endgroup& $\kbad$ & comments \\ \hline \hline
4E & $3.7858$ & $0.2641$ & $1-1.4$ & $1.224$ &  see Figures~\ref{fig:choptuik_large},~\ref{fig:choptuik_small} \\ \hline
4H$\alpha$ & $1-1.5$ & $0.66-1$ & $1-1.5$ & $1.50$ & $\kappa^*$ likely inside failure region  \\
4H$\beta$ & $1 - 1.55$ & $0.64 - 1$ & $1-1.55$ & $1.4$ &  $\kappa^*$ inside failure region  \\
4H$\gamma$ & $2.293$  & $0.436$ & $1-1.5$ & $1.32$ & see Figure~\ref{fig:4hgamma}\\ \hline
5E$\alpha$ & $1.186$ & $0.843$ & $2-2.25$ & $2.21$ & \\
5E$\beta$ & $1.665$ & $0.601$ & $2-2.25$ & $2.20$ & \\
5E$\gamma$ & $2.682$  & $0.372$ & $2-2.25$ & $2.17$ & see Figure~\ref{fig:5egamma}\\ \hline
5H$\alpha$ & $1.546$ & $0.647$ & $2.1-2.3$ & $2.43$ &    \\ 
5H$\beta$ & $1.305$ & $0.766$ & $2-2.29$ & $2.29$ &  \\
5H$\gamma$ & $1.125$ & $0.889$ & $2-2.24$ & $2.24$ &  \\
5H$\delta$ & $0.752$    &  $1.331$ & $2-2.23$  & $2.23$  & $\kappa^*$ below $\kappa=1$ gauge mode  \\
\hline
\end{tabular}
\endgroup
\caption{Results for the Choptuik exponents of the self-similar solutions.}
\label{tab:results}
\end{table}
\end{center}
\begin{figure}
    \centering
    \includegraphics[width=4in]{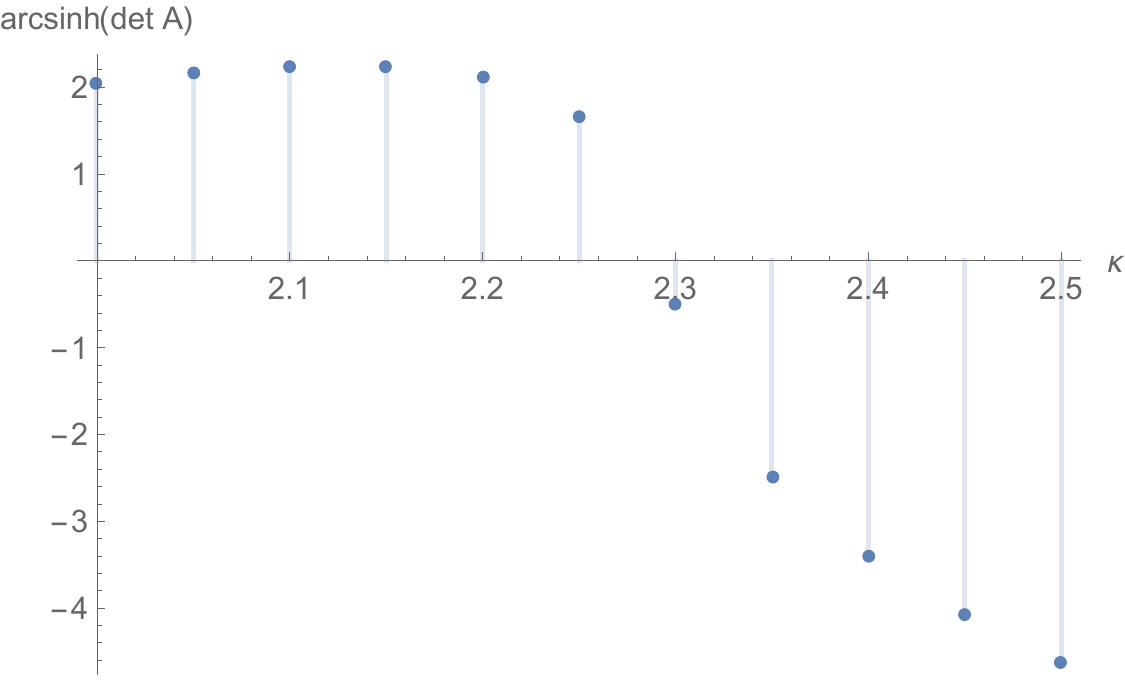}
    \caption{$\det A(\kappa)$ plot for the 4H$\gamma$ solution near the last crossing.}
    \label{fig:4hgamma}
\end{figure}
\begin{figure}
    \centering
    \includegraphics[width=4in]{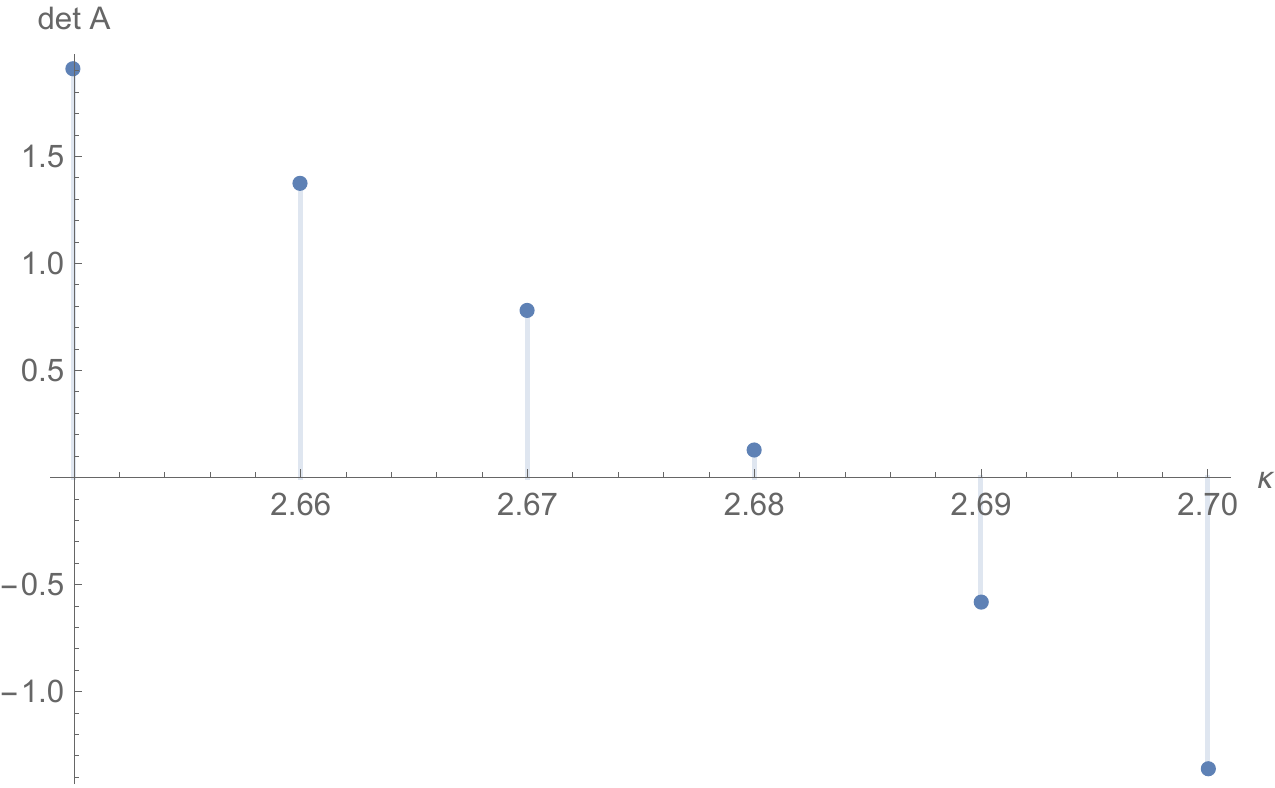}
    \caption{$\det A(\kappa)$ plot for the 5E$\gamma$ solution near the last crossing.}
    \label{fig:5egamma}
\end{figure}

\newpage

\subsection{Conclusions}

In this paper, we have outlined a numerical set up for perturbing self-similar collapse solutions to the Einstein-axion-dilaton system in arbitrary spacetime dimensions. In all cases we focus on the most relevant mode, which determines the Choptuik critical exponent. Not only does the method work for different conjugacy classes of the $\sltr$ transformation, but it can potentially encompass more general choices of matter content. In terms of results, what is perhaps most striking is that \emph{not only do the Choptuik exponents depend on the matter content (the axion-dilaton sector, in this case) and on the choice of spacetime dimension, but they also differ sizably for the various solutions of self-critical collapse that were recently found in~\cite{ours}}. 

Therefore, our results provide some clear evidence that the original expectation that Choptuik phenomena exhibit some form of universality~\cite{MA} is not fulfilled. The current analysis does not exclude that some universal behavior lies hidden in combinations of the critical exponents and other parameters of the theory. Nevertheless, this analysis and previous ones provide some definite evidence against transferring standard expectations of Statistical Mechanics to the case of critical gravitational collapse. Some uncertainties, however, still remain. In particular, the last value of $\gamma$ in the table~\ref{tab:results} is sizably larger than the others and emerges from a case where the value of $k^*$ lies below the $\kappa=1$ gauge mode. We hope to return to these important issues in the future.

\section*{Acknowledgments}
We would like to thank  E. Hirschmann, L. \'Alvarez-Gaum\'e and A. Sagnotti  for their useful comments at various stages of this work. Special thanks to A. Sagnotti for his valuable comments and critical suggestions on the final version of the paper. This work is supported by INFN (ISCSN4-GSS-PI), by Scuola Normale Superiore, and by the MIUR-PRIN contract 2017CC72MK\_003.


\end{document}